\documentclass{elsart}
\journal{Computer Physics Communications}
\usepackage{graphicx}
\usepackage{amssymb}
\usepackage{multirow}
\begin{document}
\newcommand{\wannier}{\texttt{wannier90}}
\newcommand{\pwscf}{\textsc{pwscf}}
\newcommand{\QE}{\textsc{quantum-espresso}}
\newcommand{\mmn}{$M_{mn}^{(\mathbf{k,b})}$}
\newcommand{\amn}{$A_{mn}^{(\mathbf{k})}$}
\newcommand{\umn}{$U_{mn}^{(\mathbf{k})}$}
\newcommand{\psink}{$\psi_{n\mathbf{k}}(\mathbf{r})$}
\newcommand{\rmi}{\mathrm{i}}
\newcommand{\rme}{\mathrm{e}}
\newcommand{\rmd}{\mathrm{d}}
\newcommand{\kbf}{\mathbf{k}}
\newcommand{\Gbf}{\mathbf{\Gamma}}
\newcommand{\rbf}{\mathbf{r}}
\newcommand{\R}{\mathbf{R}}
\newcommand{\bbf}{\mathbf{b}}
\newcommand{\U}{\mathbf{U}}
\newcommand{\M}{\mathbf{M}}
\newcommand{\G}{\mathbf{G}}
\newcommand{\Nkp}{N_{\mathrm{kp}}}
\newcommand{\omi}{\Omega_{\mathrm{I}}}
\newcommand{\omt}{\widetilde{\Omega}}
\newcommand{\omd}{\Omega_{\mathrm{D}}}
\newcommand{\omod}{\Omega_{\mathrm{OD}}}
\newcommand{\opt}{\mathrm{opt}}

\runauthor{Mostofi, Yates, Lee, Souza, Vanderbilt and Marzari}
\begin{frontmatter}
\title{{\wannier}: A Tool for Obtaining Maximally-Localised
  Wannier Functions}
\author[MIT]{Arash A. Mostofi\corauthref{cor}}
\corauth[cor]{Corresponding author.}
\ead{mostofi@mit.edu}
\author[LBL,Berkeley]{Jonathan R. Yates}
\author[MIT]{Young-Su Lee}
\author[LBL,Berkeley]{Ivo Souza}
\author[Rutgers]{David Vanderbilt}
\author[MIT]{Nicola Marzari}
\address[MIT]{Department of Materials Science and Engineering,
  Massachusetts Institute of Technology, Cambridge MA 02139, USA}
\address[LBL]{Materials Science Division, Lawrence Berkeley
  National Laboratory, Berkeley, CA 94720, USA} 
\address[Berkeley]{Department of Physics, University of California,
  Berkeley, CA 94720, USA} 
\address[Rutgers]{Department of Physics and Astronomy, Rutgers University,
        Piscataway, NJ 08854-8019, USA}
\begin{abstract}
We present \wannier, a program for calculating maximally-localised
Wannier functions (MLWF) from a set of Bloch energy bands that
may or may not be attached to or mixed with other bands. 
The formalism works by minimising the total spread of the MLWF in real
space. This done in the space of unitary matrices that describe
rotations of the Bloch bands at each k-point. 
As a result, \wannier\ is independent of the basis set used in the
underlying calculation to obtain the Bloch states. Therefore, it may 
be interfaced straightforwardly to any electronic structure code.
The locality of MLWF can be exploited to compute band-structure, density
of states and Fermi surfaces at modest computational
cost. Furthermore, \wannier\ is able to output MLWF for visualisation
and other post-processing purposes.
Wannier functions are already used in a wide variety of
applications. These include analysis of chemical bonding in real
space; calculation of dielectric properties via the 
modern theory of polarisation; and as an accurate and minimal basis
set in the construction of model Hamiltonians for
large-scale systems, in linear-scaling quantum Monte Carlo
calculations, and for efficient computation of material properties,
such as the anomalous Hall coefficient.
\wannier\ is freely available under the GNU General Public
License from http://www.wannier.org/.
\end{abstract}

\end{frontmatter}

%%\pagewiselinenumbers

\section*{PROGRAM SUMMARY}

\textit{Title of program:} \wannier

\textit{Catalogue identifier:}

\textit{Program Summary URL:} 

\textit{Program obtainable from:} CPC Program Library, Queen's
University of Belfast, N.~Ireland, or from the web-page http://www.wannier.org/

\textit{Licensing provisions:} This program is distributed under then
GNU General Public License v2.0 (see http://www.gnu.org/ for details)

\textit{Computers for which the program has been designed and others
  on which it has been operable:} any architecture with a Fortran 90
compiler 

\textit{Operating systems under which the program has been tested:}
Linux, Windows, Solaris, AIX, Tru64 Unix, OSX

\textit{Programming Languages used:} Fortran 90, perl 

\textit{Libraries required:} 
\begin{itemize}
\item BLAS (http://www/netlib.org/blas)
\item LAPACK (http://www.netlib.org/lapack)
\end{itemize}

Both available under open-source licenses. 

\textit{Memory used to execute with typical data:} 10 Mb 

\textit{CPU time required to execute test cases:} 1 min 

\textit{No.of bits in a word:} 32 or 64 

\textit{Has the code been vectorised or parallelised?:} No 

\textit{Number of bytes in distributed program including test data,
  etc.:} 4 600 000 

\textit{Distribution format:} gzipped tar 

\textit{Keywords:} Electronic structure, density-functional theory,
Wannier functions 

\textit{Nature of physical problem:} Obtaining maximally-localised
Wannier functions from a set of Bloch energy bands that may or may
not be entangled. 

\textit{Method of solution:} In the case of entangled bands, the
optimally-connected subspace of interest is determined by minimising a
functional which measures the subspace dispersion across the Brillouin
zone. The maximally-localised Wannier functions within this subspace
are obtained by subsequent minimisation of a functional
that represents the total spread of the Wannier functions in real
space. For the case of isolated energy bands only the second step of
the procedure is required.

%\textit{Restrictions on the complexity of the problem:} None (!)

\textit{Unusual features of the program:} Simple and user-friendly
input system. Wannier functions and interpolated band structure output
in a variety of file formats for visualisation. 

\textit{References:}
\begin{enumerate}
\item[(a)] N. Marzari and D. Vanderbilt, ``Maximally localized generalized
  Wannier functions for composite energy bands'',
  \textit{Phys. Rev. B} \textbf{56}, 12847 (1997)
\item[(b)] I. Souza, N. Marzari and D. Vanderbilt, ``Maximally localized
  Wannier functions for entangled energy bands'',
  \textit{Phys. Rev. B} \textbf{65}, 035109 (2001)
\end{enumerate}

\section*{LONG WRITE-UP}

%%%%%%%%%%%%%%%%%%%%%%
\section{Introduction}
%%%%%%%%%%%%%%%%%%%%%%

Within the independent-particle approximation, the electronic
ground state of a periodic system may be solved in terms of
a set of extended Bloch states \psink. These 
states are characterised by the good quantum numbers $n$ and
$\kbf$, which refer to the band index and crystal momentum,
respectively. Although this choice is widely used in electronic
structure calculations, alternative representations are available.
For example, the Wannier representation constitutes a description in
terms of localised functions labeled by $\R$, the lattice vector of the
cell in which the function is localised, and a band-like index $n$.

Wannier functions give a real-space picture of the electronic
structure of a system. They provide insight into the nature of the
chemical bonding and can be a powerful tool in the study of
dielectric properties via the modern theory of polarisation. 

The phase indeterminacy $\rme^{\rmi\phi_{n}(\kbf)}$ of an
isolated Bloch state \psink\ at each wave-vector $\kbf$ results
in the Wannier functions being non-unique. For a group of isolated
bands, such as the valence states of an insulator, this indeterminacy
is more general as they may undergo arbitrary unitary transformations 
\umn\ amongst themselves at each
$\kbf$. Two of the authors (NM and DV) developed a procedure
that iteratively refines these degrees of freedom such that they lead
to Wannier functions that are well-defined and localised (in the sense
that they minimise the second moment around their
centres)~\cite{marzari97:_maxim_wannier}. We  
refer to this procedure as the Marzari-Vanderbilt (MV) scheme and the 
resulting Wannier functions as maximally-localised Wannier functions (MLWF).
 
The MV method was designed to be applicable to
isolated sub-groups of bands, i.e., a group of bands that, though they
may have degeneracies with each other at certain points, are separated
from all other bands by finite gaps throughout the Brillouin zone. An
important example is the set of valence bands of an insulator. In many
applications, however, the bands of interest are not isolated and one
is interested, for instance, in the partially filled bands of a metal
close to the Fermi level. In this case the bands of interest lie
within a limited energy range but are attached to, or cross with,
other bands that extend further out in energy. Such bands are referred
to as \textit{entangled} bands.

The complication associated with treating entangled bands is that the
states of interest hybridise with unwanted bands. As a result, the
number of bands at each k-point in the relevant 
energy range may be greater than the number of Wannier functions $N$
that are required.
Before the MV localisation procedure may be applied, a prescription
for extracting $N$ bands at each k-point from the
entangled manifold of states is required.  Three of the authors 
developed just such a \textit{disentanglement} procedure, 
allowing MLWF to be determined from a set of
entangled bands~\cite{souza01:_maxim_wannier}. This
Souza-Marzari-Vanderbilt (SMV) strategy breaks 
down the procedure into two steps: first, the correct
$N$-dimensional subspace of bands at every $\kbf$ is selected;
and second, $N$ MLWF are localised from this subspace, exactly
in the same way as for an isolated set of bands.

\wannier\ is a tool for obtaining MLWF
from a set of (possibly entangled) energy bands using the
methods of MV and SMV. The principal ingredient that is required from
an electronic structure calculation is the overlap matrix between the
periodic part $\left|u_{n\kbf}\right\rangle$ of Bloch states
at neighbouring k-points (described in more detail in
Section~\ref{sec:theory}). This matrix is small and independent of the
basis used in the underlying calculation to obtain the Bloch
states. As a result \wannier\ may be interfaced to any electronic
structure code. At the time of writing, \wannier\ is able to:

\begin{enumerate}
\item[1.] Disentangle an optimally-connected $N$-dimensional
  subspace of bands at each k-point and obtain the unitary
  transformations that generate MLWF
  from a given set of bands;
\item[2.] Output MLWF in a number of
  formats suitable for visualisation;
\item[3.] Generate interpolated band structures, densities of states and
  Fermi surfaces and output them in formats suitable for
  visualisation;
\item[4.] Output matrix elements of the Hamiltonian operator in the
  MLWF basis.
\end{enumerate}

It is worth recording here a brief historical timeline of \wannier:

\begin{itemize}
\item {\bf 1997:} Two of the authors (NM and DV) develop and implement 
localisation algorithms in Fortran77; interface to the
all-bands ensemble-DFT {\sc castep}
code~\cite{marzari97:_ensem_densit_funct_theor_ab};
\item {\bf 1998:} Algorithms implemented in the {\sc cpmd}
  code~\cite{cpmd} by Silvestrelli~\cite{PhysRevB.59.9703};
\item {\bf 2001:} Three of the authors (IS, NM and DV) develop the SMV
disentanglement extension; interface to a
full-potential-linearized-augmented-plane-wave (FLAPW)
code~\cite{mno-posternak} and to any generic electronic-structure code
quickly followed;
\item {\bf 2006:} Two of the authors (AAM and JRY) completely rewrite
  the code in Fortran90, employing modern programming techniques and
  adding new functionality.
\end{itemize}

One significant benefit of the 2006 rewrite has been that in
most cases \wannier\ is faster than the original Fortran77 version by
an order of magnitude, and in some by more than two. Another
is that now the source code is easy to follow, which makes interfacing
it to electronic structure codes more straightforward. Indeed, the
first such example is already in place within the \QE\
package~\cite{pwscf}.

The formalism has seen many and diverse applications: linear-scaling
quantum Monte Carlo~\cite{PhysRevLett.87.246406}, photonic
crystals~\cite{PhysRevB.67.085204,SMHB05},
metal-insulator interfaces~\cite{stengel07}, as an efficient
interpolator for the anomalous Hall effect~\cite{xinjie06} 
and electron-phonon couplings~\cite{giustino07}, and a powerful tool
for the study of large-scale systems~\cite{lee05,mostofi-dna},
to cite only a few. In addition, MLWF are playing an increasing role in
bridging density-functional approaches and strongly-correlated ones,
to derive model Hamiltonians or as a starting point for LDA+U or
LDA+DMFT~\cite{PhysRevLett.89.167204,anisimov:075125,lechermann:125120}.
They are also closely related to muffin-tin
orbitals~\cite{PhysRevB.62.R16219,ChemPhysChem.6.1934.2005}.

The remainder of this paper is organised as follows. In
Section~\ref{sec:theory} we review briefly the theoretical background
to the problem of obtaining MLWF and a recent theoretical development
in using them to interpolate band structures on arbitrarily fine
k-point meshes. In Section~\ref{sec:technical} we give an
overview of some of the more technical aspects of the \wannier\
code. In Section~\ref{sec:structure} we describe the overall structure
of the code. We demonstrate how to install and run \wannier\ in
Sections~\ref{sec:install} and \ref{sec:run},
respectively. Finally, in Section~\ref{sec:examples} we provide a
number of example calculations and applications of the code.

%%%%%%%%%%%%%%%%%%%%%%%%%%%%%%%%
\section{Theoretical Background} \label{sec:theory}
%%%%%%%%%%%%%%%%%%%%%%%%%%%%%%%%

For an isolated set of $N$ Bloch bands \psink, a set of $N$ Wannier
functions $w_{n\R}(\rbf)=w_{n}(\mathbf{r-R})$,
$n\in[1,N]$, labelled by Bravais lattice vectors $\R$, may be
constructed as
\begin{equation}
|w_{n\R}\rangle =
\frac{V}{(2\pi)^{3}}\int_{\mathrm{BZ}} \left[ \sum_{m=1}^{N}
  U_{mn}^{(\kbf)} | \psi_{m\kbf}\rangle\right]
\rme^{-\rmi\kbf\cdot\R} \rmd \kbf ,
\end{equation}
where $\U^{(\kbf)}$ is a unitary matrix that mixes the
bands at wave-vector $\kbf$, and the Brillouin zone (BZ) integral may
also be interpreted as a unitary transformation. Different choices for
$\U^{(\kbf)}$ give rise to different Wannier functions,
with different spatial spreads, that always describe the same
manifold. The MV strategy  consists of choosing
the $\U^{(\kbf)}$ that minimise the sum of the
quadratic spreads of the Wannier functions about their centres:
\begin{eqnarray}
\Omega 
& = & \sum_{n}^{N} \left\langle \left( \rbf -
\overline{\rbf}_{n} \right)^{2} \right\rangle_{n} \nonumber\\ 
& = & \sum_{n}^{N} \left\langle r^{2} -
2\rbf\cdot\overline{\rbf}_{n} + |\overline{\rbf}_{n}|^{2}
\right\rangle_{n} \nonumber \\ 
& = & \sum_{n}^{N} \left[ \langle r^{2} \rangle_{n} - 
|\overline{\rbf}_{n}|^{2} \right]
\end{eqnarray}
where $ \langle\hat{\mathcal{O}}\rangle_{n} \equiv
\overline{\mathcal{O}}_{n} \equiv  \langle
w_{n\mathbf{0}}|\hat{\mathcal{O}}|w_{n\mathbf{0}}\rangle $. 

It is worth noting that the spread functional may be decomposed into two
contributions
\begin{equation}
\Omega=\omi + \omt, 
\end{equation}
where 
\begin{equation}
\omi=\sum_{n} \left[ 
      \langle w_{n\mathbf{0}}| r^{2} | w_{n\mathbf{0}}\rangle - 
      \sum_{m\R} \left| \langle w_{m\R}| \rbf |
      w_{n\mathbf{0}} \rangle \right|^{2} \right]
\end{equation}
and 
\begin{equation}
\omt = \sum_{n}\sum_{m\R\ne n\mathbf{0}} |\langle
w_{m\R} | \rbf | w_{n\mathbf{0}} \rangle |^{2} .
\end{equation}
It can be shown that each of these quantities is non-negative and that
$\omi$ is \textit{gauge invariant}, i.e., insensitive to changes in
the matrices $U_{mn}^{(\kbf)}$~\cite{marzari97:_maxim_wannier}.
Therefore, the minimisation of the spread functional for an isolated
set of bands just corresponds to minimising $\omt$.

%----------------------------------------
\subsection{Reciprocal-Space Formulation}\label{sec:recip}
%----------------------------------------
As shown by Blount~\cite{blount62}, matrix elements of the position
operator between Wannier functions may be expressed in
reciprocal-space as
\begin{equation}
\langle w_{n\R}|\rbf|w_{m\mathbf{0}}\rangle =
\rmi\frac{V}{(2\pi)^{3}}\int\rme^{\rmi\kbf\cdot\R} 
\langle u_{n\kbf}|\nabla_{\kbf}|u_{m\kbf}
\rangle\,\rmd \kbf  \label{eq:r-exp}
\end{equation}
and
\begin{equation}
\langle w_{n\R}|r^{2}|w_{m\mathbf{0}}\rangle =
-\frac{V}{(2\pi)^{3}}\int\rme^{\rmi\kbf\cdot\R}
\langle u_{n\kbf}|\nabla_{\kbf}^{2}|u_{m\kbf}
\rangle\,\rmd \kbf, \label{eq:r2-exp}
\end{equation}
where, as usual, the periodic part of the Bloch function is defined as
$u_{n\kbf}(\rbf)=\rme^{-\rmi\kbf\cdot\rbf}\psi_{n\kbf}(\rbf)$.
These expressions enable us to express the spread functional in terms
of matrix elements of $\nabla_{\kbf}$ and
$\nabla_{\kbf}^{2}$. 

It is assumed that the Brillouin zone is
discretised on a uniform Monkhorst-Pack mesh~\cite{monkhorst} on which
the Bloch orbitals are determined. Thus, the gradient and Laplacian
operators may be approximated by finite-difference formulas on the
$\kbf$-space mesh. If $f(\kbf)$ is a smooth function of $\kbf$, then 
\begin{equation}\label{eqn:grad-f}
\nabla_{\kbf} f(\kbf) = \sum_{\bbf}\omega_{b}\bbf\left[ f(\kbf+\bbf) -
  f(\kbf) \right]
\end{equation}
and
\begin{equation}
\langle f(\kbf) | \nabla_{\kbf}^{2} | f(\kbf) \rangle =
  \left|\nabla_{\kbf} f(\kbf)\right|^{2} = \sum_{\bbf}w_{b}\left[ 
  f(\kbf+\bbf) - f(\kbf) \right]^{2} ,
\end{equation}
where $\{\bbf\}$ are vectors connecting mesh-point $\kbf$ to
its nearest neighbours and $\omega_{b}$ is a weight factor associated
with each shell of neighbours $b=|\bbf|$. The choice of
$\bbf$-vectors and weights $\omega_{b}$ is discussed in further
detail in Section~\ref{sec:B1}.

It is worth noting that a finite k-point mesh implies an
approximation to both the self-consistent ground state that is
obtained using that mesh \emph{and} to the accuracy of the
finite-difference representation of the operators $\nabla_{\kbf}$ and
$\nabla_{\kbf}^{2}$. In principle, for a given mesh of k-points,
the latter may be improved by using higher-order finite-difference
expressions.

Having discretised Eqns.~\ref{eq:r-exp} and \ref{eq:r2-exp} in
reciprocal space, the only information required for computing the
spread functional is the overlap matrix
\begin{equation}
M_{mn}^{(\kbf,\bbf)} = 
\langle u_{m\kbf}|u_{n,\kbf+\bbf}\rangle.
\end{equation}
After some algebra, the two components of the spread functional may be
expressed as
\begin{equation}
\omi =
\frac{1}{\Nkp}\sum_{\kbf,\bbf}\omega_{b}
\sum_{m=1}^{N}\left[ 1 - \sum_{n=1}^{N} |M_{mn}^{(\kbf,\bbf)}|^{2}
  \right]
\label{eq:omegainv}
\end{equation}
and
\begin{equation}
\omt =
\frac{1}{\Nkp}\sum_{\kbf,\bbf}\omega_{b}
\left[ 
\sum_{n=1}^{N} (-\mathrm{Im}\:\ln M_{nn}^{(\kbf,\bbf)} -
\bbf\cdot\bar{\rbf}_{n})^{2}
+ \sum_{m\ne n}^{N} |M_{mn}^{(\kbf,\bbf)}|^{2}
\right],
\end{equation}
where $\Nkp$ is the number of k-points in the
Monkhorst-Pack grid
and $\bar{\rbf}_{n}$ is the centre of the $n^{\mathrm{th}}$
Wannier function, given by
\begin{equation}
\bar{\rbf}_{n} =
-\frac{1}{N_{\mathrm{kp}}}\sum_{\kbf,\bbf}
\omega_{b}\bbf\:\mathrm{Im}\:\ln M_{nn}^{(\kbf,\bbf)}.
\label{eq:wfcentre}
\end{equation}
Using these expressions, the gradient of the spread functional with
respect to infinitesimal unitary rotations of the \psink\ may be
calculated as a function of \mmn. It is then straightforward to evolve
\umn\ (and consequently \mmn) towards the solution of maximum
localisation using a steepest-descents or conjugate-gradient
minimisation algorithm.

It is worth noting that $\omt$ may be further decomposed into
band-diagonal and band-off-diagonal parts
\begin{equation}
\omt = \omd + \omod,
\end{equation}
where
\begin{eqnarray}
\omd &=& \sum_{n}\sum_{\R\ne 0} |\langle w_{n\R} | \rbf |
                      w_{n\mathbf{0}} \rangle |^{2} \\
      & = &  \frac{1}{\Nkp}\sum_{\kbf,\bbf}\omega_{b} 
\sum_{n=1}^{N} (-\mathrm{Im}\:\ln M_{nn}^{(\kbf,\bbf)} -
\bbf\cdot\bar{\rbf}_{n})^{2}, \\
 \omod &=&  \sum_{m\ne n}\sum_{\R} |\langle w_{m\R} | \rbf |
                      w_{n\mathbf{0}} \rangle |^{2} \\
      & = &  \frac{1}{\Nkp}\sum_{\kbf,\bbf}\omega_{b}
\sum_{n=1}^{N} \sum_{m\ne n}^{N} |M_{mn}^{(\kbf,\bbf)}|^{2}.
\end{eqnarray}

%---------------------------------------
\subsection{The Case of Entangled Bands} \label{sec:SMV}
%---------------------------------------

The above description is sufficient to obtain MLWF for an
isolated set of bands, such as the valence states in an insulator.
In order to obtain MLWF for entangled
energy bands, e.g., for metallic systems or the conduction bands of an
insulator, we use the ``disentanglement'' procedure introduced by
SMV~\cite{souza01:_maxim_wannier}.

The strategy proceeds as follows. An energy window (the ``outer
window'') is defined by the user such that at each k-point
there are $N^{(\kbf)}_{\mathrm{win}}\ge N$ states within the
window. An orthonormal set of $N$ Bloch states is obtained at each
k-point, defining an $N$-dimensional subspace
$\mathcal{S}(\kbf)$, by performing a unitary transformation amongst
the $N^{(\kbf)}_{\mathrm{win}}$ states that fall within the energy
window:
 \begin{equation}
| u_{n\kbf}^{\opt}\rangle = \sum_{m\in
  N^{(\kbf)}_{\mathrm{win}}} U^{\mathrm{dis}(\kbf)}_{mn} |
  u_{m\kbf} \rangle ,
\end{equation}
where $\U^{\mathrm{dis}(\kbf)}$ is a rectangular
 $N^{(\kbf)}_{\mathrm{win}}\times N$ matrix.\footnote{As
    $\U^{\mathrm{dis}(\kbf)}$ is rectangular, this is a
 unitary operation in the sense that
 $(\U^{\mathrm{dis}(\kbf)})^{\dagger}
 \U^{\mathrm{dis}(\kbf)}=\mathbb{I}$.}

Recall that $\omi$ is invariant under gauge
 transformations within a given subspace. Thus $\omi$
 may be considered as a functional of $\mathcal{S}(\kbf)$. The
 subspace selection proceeds by minimising $\omi$ with
 respect to the matrices
 $\U^{\mathrm{dis}(\kbf)}$~\cite{souza01:_maxim_wannier}.
For a physical interpretation of this procedure,
Eqn.~\ref{eq:omegainv} for $\omi$ may be rewritten as
\begin{equation}
\omi =
\frac{1}{\Nkp}\sum_{\kbf,\bbf}\omega_{b}
\mathrm{tr}\left[ \hat{P}_{\kbf}\hat{Q}_{\kbf+\bbf}\right],
\end{equation}
where $\hat{P}_{\kbf}=\sum_{n=1}^{N}|u_{n\kbf}\rangle\langle u_{n\kbf}|$ is
the projector onto $\mathcal{S}(\kbf)$, and
$\hat{Q}_{\kbf}=\mathrm{\bf 1}-\hat{P}_{\kbf}$. Now it can be seen that
$\omi$ is a measure of the degree of mismatch between the
subspaces $\mathcal{S}(\kbf)$ and $\mathcal{S}(\kbf+\bbf)$. Minimising
$\omi$ corresponds to choosing self-consistently at every
$\kbf$ the subspace that has maximum overlap as $\kbf$ is varied.  

 Once the projected $N$-dimensional subspace at each k-point
 has been found, the MV localisation procedure described above may be
 used to minimise $\omt$ within that subspace and 
 hence obtain MLWF. As an alternative to this two
 step minimisation, a procedure to minimise the total spread function $\Omega$
 has been proposed~\cite{thygesen:026405} and was found to give
 very similar MLWF to the present scheme.

 It may be the case that the energy bands of the projected subspace do
 not correspond to any of the original energy bands due to mixing
 between states. 
 In order to preserve exactly the properties of a
 system within a given energy range (e.g., around the Fermi level) a
 second \textit{inner} energy window is introduced. States lying
 within this inner, or frozen, energy window are included unchanged in
 the projected subspace. The reader is referred to
 Ref.~\cite{souza01:_maxim_wannier} for further details of the
 algorithm.

It is worth noting that the MLWF themselves are never explicitly
constructed unless required for
visualization or other post-processing purposes. Minimisation
of the spread functional results in finding the converged
$\U^{\mathrm{dis}(\kbf)}$ (if disentanglement was used) and
$\U^{(\kbf)}$. Along with $\M^{(\kbf,\bbf)}$, this is sufficient to
define the centres and spreads of the MLWF. If the periodic parts of
the Bloch wavefunctions are available, then the MLWF may be
calculated on a real-space grid. For systems with time-reversal symmetry,
we always find the MLWF
corresponding to the minimum spread to be real, apart from a global
complex phase factor, in empirical agreement
with a recent claim of a mathematical proof~\cite{Brouder:PRL.98.046402}.

%---------------------------------
\subsection{Wannier Interpolation}
%---------------------------------

Having found $\U^{\mathrm{dis}(\kbf)}$ and  $\U^{(\kbf)}$ for the target
system, it is straightforward to express the Hamiltonian in the basis
of MLWF. We first obtain the Hamiltonian in the basis of
rotated Bloch states \begin{equation}\label{eq:int1}
H^{(W)}(\kbf)=(U^{(\kbf)})^{\dagger}(U^{\mathrm{dis}(\kbf)})^{\dagger}
H(\kbf)U^{\mathrm{dis}(\kbf)}U^{(\kbf)}
\end{equation}
where $H_{nm}(\kbf)=\varepsilon_{n\kbf}\delta_{nm}$. Next we find its Fourier
sum
\begin{equation}
H_{nm}(\R)=\frac{1}{N_0}\sum_{\kbf} e^{-\rmi\kbf\cdot\R}H^{(W)}_{nm}(\kbf).
\end{equation}
This operation is done once and for all for each of the $N_0$ lattice
vectors $\R$ lying in a supercell conjugate to the
$\kbf$-mesh (in practice we choose a Wigner-Seitz supercell centred on
the origin~\cite{wanint}). $H_{nm}(\R)$ can be recognised as the matrix
elements of the Hamiltonian between MLWF. Due to the strong localisation
of the MLWF, the matrix elements $H_{nm}(\R)$ decay rapidly with $R$. In
the spirit of a Slater-Koster interpolation scheme~\cite{slater-koster} this
allows us to compute the Hamiltonian on a much finer mesh of k-points in
the original Bloch space. We can perform the inverse Fourier transform
\begin{equation}
H_{nm}(\kbf')=\sum_{\R} e^{\rmi \kbf'\cdot\R}H_{nm}(\R),
\end{equation}
to yield the interpolation of Eqn.~\ref{eq:int1} onto an arbitrary
k-point $\kbf'$. The final step is to diagonalise $H_{nm}(\kbf')$
to obtain the interpolated band energies. 

By construction, the interpolated band energies coincide with the true
bandstructure at the original k-points of the Monkhorst-Pack mesh (if
the SMV scheme is used, this is only guaranteed to occur within the
inner energy window). At intermediate k-points, the accuracy of the
interpolation is dependent on the density of the original
mesh~\cite{lee05,wanint}. 

Within \wannier\ this interpolation procedure can be used to
obtain plots of band structure, density of states and the Fermi
surface at modest computational cost. The Wannier interpolation
formalism is rather general and can also be used to interpolate arbitrary
periodic operators~\cite{wanint}.

%%%%%%%%%%%%%%%%%%%%%%%%%%%%%%%%
\section{Some Technical Aspects} \label{sec:technical}
%%%%%%%%%%%%%%%%%%%%%%%%%%%%%%%%

%----------------------------------------------------
\subsection{Initial guess for iterative minimisation} \label{sec:proj}
%----------------------------------------------------

The iterative minimisation of $\omi$ begins with an initial guess for
the subspaces $\mathcal{S}(\kbf)$. One possible choice is to start
from the initial, random phases of the Bloch states provided by the
ab initio code. Alternatively, we may define a set of $N$ trial
functions $g_{n}(\rbf)$, $n\in[1,N]$, as an initial approximation to
the $N$ MLWF. These are projected onto the cell-periodic parts of the
$N^{(\kbf)}_{\mathrm{win}}$ Bloch eigenstates inside the energy window:
\begin{equation}
|\phi_{n\kbf}\rangle = \sum_{m=1}^{N^{(\kbf)}_{\mathrm{win}}}
 A_{mn}^{(\kbf)}|u_{m\kbf}\rangle, 
\end{equation}
where $A_{mn}^{(\kbf)}=\langle u_{m\kbf}|g_{n}\rangle$ is an
$N^{(\kbf)}_{\mathrm{win}}\times N$ 
matrix. Orthonormalising the resulting $N$ functions
$\{|\phi_{n\kbf}\rangle\}$ via a L\"{o}wdin transformation, we find
\begin{equation}
|u_{n\kbf}^{\opt}\rangle = \sum_{m=1}^{N}
 (S^{-1/2})_{mn}|\phi_{m\kbf}\rangle = \sum_{m=1}^{N^{(\kbf)}_{\mathrm{win}}}
 (AS^{-1/2})_{mn}|u_{m\kbf}\rangle,
\end{equation}
where $S_{mn}\equiv
S^{(\kbf)}_{mn}=\langle\phi_{m\kbf}|\phi_{n\kbf}\rangle=(A^{\dagger}A)_{mn}$, 
and $\mathbf{AS}^{-1/2}$ is used as the initial guess for
$\U^{\mathrm{dis}(\kbf)}$.

The same trial orbitals $\{|g_{n}\rangle\}$ can also be used to
initialize the minimization of $\omt$. Using a similar projection
procedure to the one described above, an initial guess for the
$N\times N$ unitary matrices $\U^{(\kbf)}$ is obtained at each
k-point.

We now come to the choice of trial orbitals. As the minimisation
scheme is quite robust, one option is to choose a set of
spherically-symmetric Gaussian functions whose centres are chosen
randomly; \wannier\ supports this option.
On the other hand, a user may wish to apply chemical and physical
insight in order to select a better starting point. An ab
  initio code that is interfaced to \wannier\ may use any localised
functions desired to construct the projections \amn. For convenience,
we have defined a standard set of projection functions that should
suffice for most situations. This is the set of atomic orbitals 
of the hydrogen atom, which is a convenient choice for two reasons:
first, analytical mathematical forms exist in terms of the good
quantum numbers $n$, $l$ and $m$; and second, hybrid orbitals (sp,
sp$^{2}$, sp$^{3}$, sp$^{3}$d etc.) may be constructed by simple
linear combination $|\phi\rangle = \sum_{nlm} C_{nlm}|nlm\rangle$, for
some coefficients $C_{nlm}$. The angular parts of our projection
functions are not the canonical spherical harmonics $Y_{lm}$ of the
hydrogenic Schr\"{o}dinger equation, but rather the \textit{real} (in
the sense of non-imaginary) states $\Theta_{lm_{\mathrm{r}}}$,
$m_{\mathrm{r}} \in [1,2l+1]$, obtained by unitary transformation of
the $Y_{lm}$.

Each localised function is associated with a site and an angular
momentum state. Optionally, one may define the spatial orientation, the
radial form and the diffusivity for each function. \wannier\ is able
to project onto functions with s, p, d and f symmetry, plus the
hybrids sp, sp$^2$, sp$^3$, sp$^3$d, sp$^3$d$^2$. The user is referred
to the documentation in the \wannier\ distribution for mathematical
definitions and details on how to specify projection functions in the
input file.

In general, the spread functional $\Omega$ has local
minima and, occasionally, the minimisation becomes trapped in one. In
other words, the final solution may depend on the initial
choice for $\mathcal{S}(\kbf)$ and hence $\{g_{n}(\rbf)\}$. In most
cases, we find that these local minima give 
rise to MLWF that are complex, i.e., they have significant imaginary
parts. 
In some cases, MLWF associated
with local minima of $\Omega$ \textit{are} found to be real and the
reader is referred to Ref.~\cite{Si_MLWFS_2007} for more details.

%-------------------------------------------
\subsection{Choosing the {\bf b} vectors} \label{sec:B1}
%-------------------------------------------

As discussed in Section \ref{sec:recip}, in order to compute the
spread functional we require a finite-difference formula for
$\nabla_{\bf k}$. We now describe an automatic procedure to choose,
for cells of arbitrary symmetry, the simplest such formula.

It is assumed that the Brillouin zone is
discretised on a uniform Monkhorst-Pack mesh~\cite{monkhorst}.
The vectors $\{\bbf\}$ connect each mesh-point $\kbf$ to
its nearest neighbours. $N_{\mathrm{sh}}$ shells of
neighbours are included in the finite-difference formula, with
$M_{s}$ vectors in the $s^{\mathrm{th}}$ shell. For
$\nabla_{\bf k}$ to be correct to linear order we require that (see
Eqn. B1 of Ref.~\cite{marzari97:_maxim_wannier})
\begin{equation}\label{eq:b1}
\sum_{s}^{N_{\mathrm{sh}}} w_s \sum_{i}^{M_s}
b^{i,s}_{\alpha}b^{i,s}_{\beta}=\delta_{\alpha\beta}
\end{equation}
where ${\bf b}^{i,s}$, $i \in [1,M_{s}]$, is the
$i^{\mathrm{th}}$ vector belonging to the $s^{\mathrm{th}}$ shell with
associated weight $w_s$, and $\alpha$ and $\beta$ run over the three
Cartesian indices. If $f({\bf k})$ is a smooth function of ${\bf k}$,
then its gradient may be expressed according to Eqn.~\ref{eqn:grad-f}.
For the case of a linear function $f({\bf k})=f_0 + {\bf g}\cdot{\bf
  k}$, this finite-difference formula gives the exact result $\nabla_{{\bf
  k},\alpha}=\sum_{\bf b} w_b b_{\alpha}
  g_{\beta}b_{\beta}=\delta_{\alpha\beta}g_{\beta}=g_{\alpha}$, where
  summation convention over repeated Greek indices is assumed.

In order to find the weights $w_s$, we notice that Eqn. \ref{eq:b1} is
symmetric in the Cartesian indices, therefore, there are six independent
elements that may be expressed through a matrix equation
\begin{equation}
{\bf A}{\bf w}={\bf q}, \label{eq:awq}
\end{equation}
where ${\bf q}$ is a vector of length six containing the six elements
of the lower triangular part of
$\delta_{\alpha\beta}$, ${\bf w}$ is the vector of weights with length
$N_{\mathrm{sh}}$, and 
${\bf A}$ is a 6 $\times$ $N_{\mathrm{sh}}$ matrix given by $A_{j,s}=\sum_i
b^{i,s}_{\alpha}b^{i,s}_{\beta}$, where there is a one-to-one
correspondence between $j \in [1,6]$ and the six independent
pairings of $\alpha$ and $\beta$.
${\bf A}$ may be factorised using a singular value decomposition~\cite{SVD},
\begin{equation}
{\bf A}={\bf U}{\bf D}{\bf V}^{T},
\end{equation}
which permits inversion of Eqn.~\ref{eq:awq} and solution for the
shell weights, 
\begin{equation}
{\bf w}={\bf V}{\bf D}^{-1}{\bf U}^{T}{\bf q}.
\end{equation}
Our automatic procedure for choosing the shells is as follows:
we add shells in order of increasing $|{\bf b}|$; with each additional
shell we obtain the shell weights, and check to see if Eqn. \ref{eq:b1} is
satisfied; if not, then we add a new shell and repeat the
procedure. A new shell may be linearly dependent on the
existing shells, in which case one or more of the singular values will be
close to zero and we reject the new shell. When the Bravais lattice
point group is cubic, the first shell of nearest neighbours is
sufficient. The maximum number of shells required is six (for triclinic
symmetry). For the case of a very elongated unit cell it may be
necessary for the routine to search through a large number of shells
in order to find the correct combination.

With the above finite-difference formalism, the quadratic spread
converges only polynomially with the sampling of the Brillouin
zone. This slow convergence is a property of the spread
operator, whereas the underlying MLWF converge rapidly with the k-point
density~\cite{stengel06b}. It may be possible to achieve improved
accuracy and k-point convergence by using higher-order
finite-difference formulas for $\nabla_{\bf k}$, but this has not been
explored.

In some cases, the automatic procedure outlined above might not find
the shells and weights that give the most spherically symmetric
representation of the position operator. When this happens, special
care may be required to ensure that the most symmetric choice of
shells is made~\cite{mno-posternak}, and this can be done explicitly
(i.e., by hand) in the input file.

%--------------------------------------------------
\subsection{Specific Algorithms for $\Gamma$-Point Sampling} \label{sec:gamma}
%--------------------------------------------------

For isolated molecules or extended systems of large unit cell,
single $\Gamma$-point sampling in the reciprocal space 
can provide an accurate description of the physical quantities of interest.
When the $\Gamma$-point is sampled exclusively, the
computational cost can be reduced by exploiting the symmetries of the 
overlap matrices $\M^{(\Gbf,\bbf)}$.

First, at $\Gamma$, 
the ${\bf b}$-vectors are linear combinations of the primitive vectors
of the reciprocal lattice, 
thus imposing the following conditions: 
\begin{equation}
\psi_{n,\Gbf+\bbf}(\rbf) = \psi_{n\Gbf}(\rbf) \quad , \quad
u_{n,\Gbf+\bbf}(\rbf) = \rme^{-\rmi\bbf\cdot\rbf}u_{n\Gbf}(\rbf).
\end{equation}
Then, it follows that
$\M^{(\Gbf,-\bbf)}$ is the
Hermitian conjugate of $\M^{(\Gbf,\bbf)}$,
\begin{equation}
\begin{array}{rcl}
M_{mn}^{(\Gbf,\bbf)} &=& \langle u_{m\Gbf}|u_{n,\Gbf+\bbf}\rangle       
=\langle u_{n,\Gbf+\bbf}|u_{m\Gbf}\rangle^\ast
=\langle u_{n\Gbf}|u_{m,\Gbf-\bbf}\rangle^\ast \\
&=&\left( M_{nm}^{(\Gbf,-\bbf)} \right )^\ast.
\end{array}
\label{eq:trans}
\end{equation}
Second, at $\Gamma$, the Bloch eigenfunctions may be chosen to be real.
Then, the overlap matrices become symmetric, $M_{mn}^{(\Gbf,\bbf)} =
M_{nm}^{(\Gbf,\bbf)}$, and only the upper or the lower half of
$\M^{(\Gbf,\bbf)}$ is independent.

\wannier\ includes a $\Gamma$-only branch of algorithms that is able
to exploit these symmetries.  
These algorithms may be activated by means of a logical keyword in the
input file and 
they rely on the above relations being satisfied, i.e., the  
Bloch wavefunctions must be real.

There are several other advantages beside
the significant decrease in the number of operations.
In the disentanglement procedure~\cite{souza01:_maxim_wannier},
the diagonalisation of a complex Hermitian matrix at each iteration step
is replaced by that of a real symmetric matrix, 
which ensures that $\U^{\mathrm{dis}(\Gbf)}$ is real.
The localisation procedure in the $\Gamma$-only branch adopts
an efficient algorithm proposed by Gygi \emph{et al.}~\cite{gygi03}.
This method minimises $\omod$ (not $\omt$)
by simultaneously diagonalizing
a set of real symmetric matrices, $\{ \mathrm{Re}\, \M^{(\Gbf,\bbf)}, 
\mathrm{Im}\, \M^{(\Gbf,\bbf)} \}$.
The $\U^{(\Gbf)}$ from this method is inherently real,
giving real MLWF as a final product. 

It is worth mentioning that
these MLWF are not necessarily identical to those obtained from the
MV localisation procedure
unless $\omd$
is strictly zero due to symmetry of the system.

%--------------------------------------------------
\subsection{Representative Timings and Convergence} \label{sec:timings}
%--------------------------------------------------

 In this section, timings and convergence characteristics for
 representative calculations are  presented. 
 The initial band-structure calculations are performed using the \pwscf\
 code~\cite{pwscf}.
 The self-consistent ground state is obtained, then the required
 overlap matrices and projections are calculated using the
 post-processing routine \texttt{pw2wannier90}, supplied with the
 \pwscf\ distribution. \wannier\ is then used to obtain the MLWF. 
 The \pwscf\ and \texttt{pw2wannier90}
 calculations were performed on a four-processor (dual dual-core)
 Intel Woodcrest 5130 2.0GHz desktop computer and the 
 \wannier\ calculations on a single processor of the same machine. The
 gauge-invariant and gauge-dependent spreads were converged to a
 tolerance of $10^{-10}$\,\AA$^{2}$.

\subsubsection{Crystalline Silicon}

 First, we consider a two-atom unit cell of crystalline silicon. A
 kinetic-energy cutoff of 25\,Ry is used for the plane-wave 
 expansion of the valence wavefunctions. 
 The
 core-valence interaction is
 described by means of norm-conserving pseudopotentials in separable
 Kleinman-Bylander form~\cite{kleinman1}.
Table~\ref{tab:wan90-timings} gives
 timings for obtaining the valence MLWF for various densities of
 Monkhorst-Pack k-point grid. Bond-centred s-type functions were
 used for the initial projection. It can be seen that the most time
 consuming part of the procedure is the computation of the band
 structure, followed by construction of \mmn\ and \amn, with \wannier\
 taking only 5-7\% of the total time to perform the localisation of
 the MLWF and to write files for their visualisation.

\begin{table}[h]
\begin{center}
%%%\begin{tabular}{|c||c|c|c|c|}
\begin{tabular}{ccccc}
\hline
$\Nkp$ & $T_{\mathrm{SCF}}$  & $T_{M,A}$
 & $T_{\mathrm{W90}}$  & $N_{\mathrm{iter}}$ \\ \hline
64   &  7  &  5  & $<1$ &  5  \\ %\hline
512  & 52  & 44  & 4    & 14  \\ %\hline
1728 & 190 & 142 & 13   & 30  \\ %\hline
4096 & 462 & 329 & 56   & 96  \\ \hline
\end{tabular}
\caption{Timings (in seconds) for the valence states of crystalline
  silicon with (i) different densities of Monkhorst-Pack mesh.
  $T_{\mathrm{SCF}}$ is the time taken by \pwscf\ to obtain the ground
  state wavefunctions at all $\Nkp$ k-points of the
  Monkhorst-Pack mesh, $T_{M,A}$ is the time
  taken by \texttt{pw2wannier90} to calculate \mmn\ and \amn, and
  $T_{\mathrm{W90}}$ is the time taken by \wannier\ to localise and
  plot the MLWF. $N_{\mathrm{iter}}$ is the number of iterations
  required to minimise the total spread.} 
\label{tab:wan90-timings}
\end{center}
\end{table}

\begin{figure}
\begin{center}
\scalebox{0.3}{\includegraphics{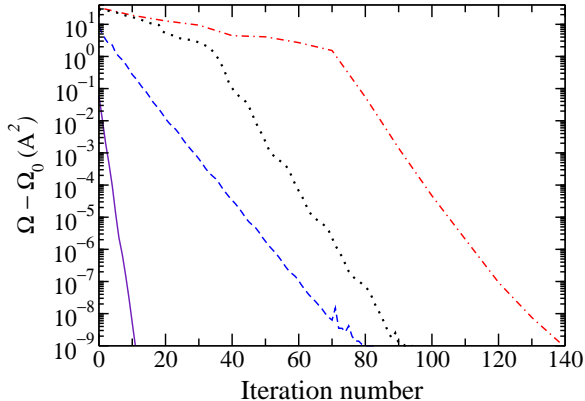}}
\caption{Convergence of total spread $\Omega$ for crystalline silicon
  with an $8\times 8\times 8$ Monkhorst-Pack k-point grid. 
  Solid curve: bond-centred s projection. 
  Dashed curve: atom-centred sp$^2$-like projection. Dotted curve: 
  atom-centred s and p projection. Dot-dashed curve: randomly
  centred spherical s-type projection. $\Omega_{0}$ is the converged
  total spread and is independent of the initial projection used.}
\label{fig:si-proj}
\end{center}
\end{figure}

In Figure~\ref{fig:si-proj} the effect of using different initial
projections for the MLWF is demonstrated. It is worth noting that
even when choosing randomly-centred s-type functions the correct MLWF
are found in less than 10\% of the total time. 

\subsubsection{Fullerene}

We consider an isolated fullerene molecule (C$_{60}$)
as a test case for the $\Gamma$-only algorithms. 
The molecule was placed in a cubic supercell of side-length
40\,a$_{0}$ (21.2\,\AA), and the calculations were performed with
$\Gamma$-point sampling, ultrasoft pseudopotentials~\cite{vanderbilt1},
and a plane-wave cut-off of 30\,Ry. 

Fullerene has 120 valence states in total.
For the localisation of the valence MLWF,
two choices of initial projections 
are compared, one being 120 randomly-centred s-orbitals
and the other being 120 s-orbitals placed on the
MLWF centres generated by the former run. The
disentanglement procedure is tested
on the extended 150 states
comprising 120 valence states
and 30 states that cover the full $\pi^\ast$-manifold.
Those 30 states are disentangled from
the 100 unoccupied eigenfunctions spanning
up to 7.5\,eV above the HOMO level.
The final set of MLWF are found to consist of
60 atom-centred p$_z$ orbitals and
90 bond-centred $\sigma$-bonding orbitals,
similar to the MLWF shown in Fig.~\ref{fig:graphite_wf}.

\begin{table}[h]
\begin{center}
%%%\begin{tabular}{|c||c||c||c|c||c|c||c|c|}
\begin{tabular}{|c|cccccccc|}
\hline
 Projection & $\Gamma$ & $N$ & $T_{M,A}$ & $T_{\mathrm{W90}}$ & 
 $T_{\mathrm{dis}}$ & $N_{\mathrm{iter}}^{\mathrm{dis}}$ & 
 $T_{\mathrm{loc}}$ & $N_{\mathrm{iter}}^{\mathrm{loc}}$ \\
 \hline 
 \multirow{2}{*}{randomly-centred s}                  
 & F & 120  &  901 & 33  & n/a & n/a & 33 & 226 \\ %\cline{2-9}
 & T & 120  &  686 & 1 & n/a & n/a &   1  & 16 \\ \hline
 \multirow{2}{*}{bond-centred s}                      
 & F & 120  &  893 & 18 & n/a & n/a & 18 & 117 \\ %\cline{2-9}
 & T & 120  &  694 & $<1$ & n/a & n/a & $<1$ & 11 \\ \hline
 bond-centred s,  & F & 150  & 1595 & 46 & 22 &  75  & 24 & 83 \\ %\cline{2-9}
 atom-centred p$_z$ & T & 150  & 1017 & 12 & 11 &  75  & 1 & 9 \\ 
\hline
\end{tabular}
\caption{Timings (in seconds) for MLWF in fullerene from
  general k-point and $\Gamma$-specific schemes.
  Different initial projections are tested.
  ``T'' in the second column ($\Gamma$) indicates that
  the $\Gamma$-only branch of algorithms is used. 
  $N$  is the number of MLWF obtained.
  $T_{M,A}$ is the time
  taken to calculate \mmn\ and \amn, and $T_{\mathrm{W90}}$ is the
  total time taken by \wannier\ to disentangle and localise MLWF. 
  The individual time and the number of iterations
  for each of these operations are
  given as $T_{\mathrm{oper}}$ and $N_{\mathrm{iter}}^{\mathrm{oper}}$
  (where $\mathrm{oper}$=$\mathrm{dis}$ or $\mathrm{loc}$), respectively.} 
\label{tab:bucky}
\end{center}
\end{table}

Timings for these calculations 
are summarized in Table~\ref{tab:bucky}.
$T_{M,A}$ decreases by more than 20\% 
when the $\Gamma$-only branch is used,
because only half the $\bbf$-vectors are needed.
The post-processing routine \texttt{pw2wannier90}
is not yet optimized to take full advantage of 
real wavefunctions, and 
we expect at least 50\% of time reduction 
in the fully optimized case.

The minimisation method used in the $\Gamma$-only branch
is apparently more efficient than the conjugate-gradient algorithm,
especially when the initial projections are
far from the final, converged MLWF.
$\omd$ vanishes in the case of a simple cubic lattice,
and $\Omega$ from the two different minimization methods 
converges to an identical value up to the given tolerance.
$T_{\mathrm{dis}}$ decreases by 50\% for the reason
discussed in  Section \ref{sec:gamma},
but $N_{\mathrm{iter}}^{\mathrm{dis}}$ 
is the same, as it should be,
and $\omi$ is identical within machine precision.

%%%%%%%%%%%%%%%%%%%%%%%%%%%%%%%%%%
\section{Structure of the Program} \label{sec:structure}
\begin{figure}
\begin{center}
\scalebox{0.54}{\includegraphics{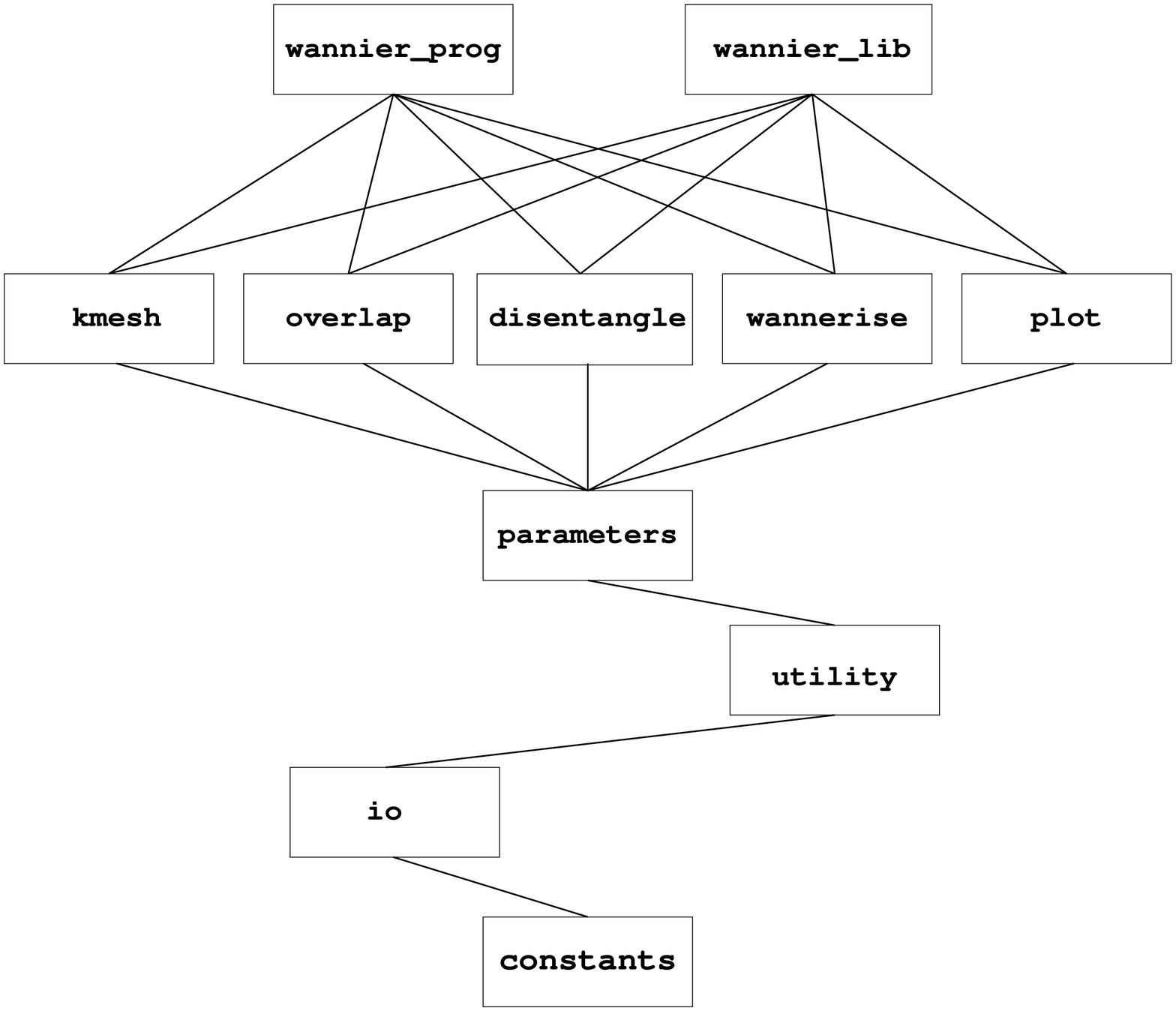}}
\caption{Schematic structure of the program} \label{fig:prog_structure}
\end{center}
\end{figure}

%%%%%%%%%%%%%%%%%%%%%%%%%%%%%%%%%
The schematic structure of the program is outlined in
Fig.~\ref{fig:prog_structure}. Each box represents a Fortran90 module
and the lines represent module dependencies. Modules only use data and
subroutines from lower modules in the diagram. A description of the
purpose of each module is given below.

\begin{itemize}
\item \texttt{constants}: definition of constants (e.g., $\pi$)
\item \texttt{io}: error handling, timing, and input and output units
\item \texttt{utility}: commonly used operations such as
  conversion of Cartesian to fractional co-ordinates, string
  functions, matrix multiplication wrapper etc.
\item \texttt{parameters}: all physical parameters relevant to
  the calculation and subroutines for reading them from the input file at the
  start of a calculation. Subroutines for writing
  checkpoint files for restarting previous runs.
\item \texttt{kmesh}: set up of the
  framework for reciprocal-space derivatives and, if in
  post-processing mode, writing a
  file which communicates to an
  ab initio code information on how to
  calculate \mmn\ and \amn
\item \texttt{overlap}: reading of the overlap matrices
  \mmn\ and \amn
\item \texttt{disentangle}: if using entangled energy bands, finding the
  optimal subspace within a specified energy window by minimising
  the gauge-invariant spread
\item \texttt{wannierise}: finding the unitary transformations
  \umn\ amongst the energy bands which give rise to MLWF
\item \texttt{plot}: routines for output of Wannier functions, Fermi
  surface and band structure in file formats suitable for visualisation
\item \texttt{wannier\_prog}: the main program
\item \texttt{wannier\_lib}: library routines. \wannier\
  may be invoked directly from within an ab initio
  code as a set of library calls. The reader is referred to
  Section~\ref{sec:wannier_lib} for a brief description and to the
  documentation in the \wannier\ distribution for full details.
\end{itemize}

%%%%%%%%%%%%%%%%%%%%%%
\section{Installation} \label{sec:install}
%%%%%%%%%%%%%%%%%%%%%%
\wannier\ is distributed as a gzipped tar file
(http://www.wannier.org/). On Linux platforms, for example, it may be
unpacked by typing
\begin{verbatim}
> tar zxvf wannier90.tar.gz
\end{verbatim}
which creates a directory containing the source files, documentation,
examples etc.
 
Compilation is straightforward. From the \texttt{config} directory of
the distribution the user should select the \texttt{make.sys.plat}
file which corresponds most closely to the platform being used and
copy it to the root directory of the distribution, renaming it
\texttt{make.sys} in the process. The values of system-dependent
parameters (e.g., the location of the BLAS and LAPACK 
libraries, the name of the Fortran compiler, the Fortran
optimisation flags etc.) that are defined therein should be modified to
correspond to the user's particular system. Once this has been done,
typing \texttt{make} from the root directory of the distribution will
create an executable \texttt{wannier90.x}. Further details and make
options may be found in the file \texttt{INSTALL.readme} in
the root directory of the distribution.

%%%%%%%%%%%%%%%%%%%%%%%%%%%%
\section{Running {\wannier}} \label{sec:run}
%%%%%%%%%%%%%%%%%%%%%%%%%%%%

Before running \wannier\ the user must perform a self-consistent
first-principles calculation on the system of interest in order to
obtain a set of Bloch energy bands from which MLWF may then be
constructed. Once the Bloch bands have 
been computed \wannier\ may be operated in a post-processing mode as
described below.  

The master input file for \wannier\ is called \texttt{seedname.win},
where \texttt{seedname} is the prefix of all of the input
and output files. The input system is designed to be simple and
user-friendly and is described comprehensively in the 
documentation that is contained within the \wannier\ distribution. An
example input file is provided in Appendix~\ref{app:testrun}.

\wannier\ must be run twice. On the first pass the command line option
\texttt{-pp} must be used, as follows: 

\begin{verbatim}
> wannier90.x -pp seedname
\end{verbatim}

This causes the code to read the master input file
\texttt{seedname.win} and generate the file \texttt{seedname.nnkp}
that contains all the information necessary to construct the overlap
matrices \mmn\ and \amn\ from the Bloch bands already obtained from
a first-principles calculation. In order to interface an ab
  initio code to \wannier\ one needs to write subroutines that 
read \texttt{seedname.nnkp}, compute the overlap matrix \mmn\ and,
optionally, \amn from the Bloch bands and write these matrices in the
appropriate format to files called \texttt{seedname.mmn} and
\texttt{seedname.amn}, respectively. If using the disentanglement
procedure or plotting a band structure, density of states or Fermi
surface, \wannier\ also requires the 
eigenvalues $\varepsilon_{n\kbf}$ corresponding to 
the Bloch states $\psi_{n\kbf}(\rbf)$, which should be
written to a file called \texttt{seedname.eig}. 
The reader is referred to 
the documentation that is contained within the \wannier\ distribution
for complete details of these files.

Once the necessary files have been written, \wannier\ must be run
again, this time without any command-line options:

\begin{verbatim}
> wannier90.x seedname
\end{verbatim}

On this pass, the code reads the overlap matrices and eigenvalues (if
required) from 
file and performs the maximal-localisation procedure as outlined in
Section~\ref{sec:theory}, writing the output to a file
\texttt{seedname.wout}.

At the time of writing, the \pwscf\ code (a part of the \QE\
package~\cite{pwscf}), which is available under GNU General Public
License, has a 
\wannier\ interface in the form of a post-processing program called
\texttt{pw2wannier90}. It is the authors' hope that \wannier\ is a
sufficiently useful tool for investigators to be motivated to write
interfaces to other electronic structure codes. 

%------------------------
\subsection{Library mode} \label{sec:wannier_lib}
%------------------------

\wannier\ may also be compiled as a library and invoked with
subroutine calls from
within an ab initio code. The command

\begin{verbatim}
> make lib
\end{verbatim}

creates a library \texttt{libwannier.a} in the root directory of the
distribution.  The library mode of \wannier\
works along exactly the same lines as the post-processing mode,
described above. The formal difference is that in the latter,
information is passed between the ab initio code and
\wannier\ via {\it files}, such as \texttt{seedname.nnkp} and
\texttt{seedname.mmn} etc., whereas, in the former, it is done via
direct {\it calls} to library subroutines. As before, there are two
stages: first, a call to the 
library subroutine \texttt{wannier\_setup}, which returns the
information necessary to construct the overlap matrices \mmn\ and
\amn\ from the Bloch bands; second, a call to the subroutine
\texttt{wannier\_run}, which takes as input the \mmn\ matrix (and
\amn\ and the eigenvalues $\varepsilon_{n\kbf}$, if
required) and performs the maximal-localisation procedure as outlined
in Section~\ref{sec:theory}. The reader is referred to the
documentation in the \wannier\ distribution for further details.

%%%%%%%%%%%%%%%%%%
\section{Examples} \label{sec:examples}  %JRY
%%%%%%%%%%%%%%%%%%

\subsection{Graphite}
\begin{figure}[h]
\begin{center}
\scalebox{0.35}{\includegraphics{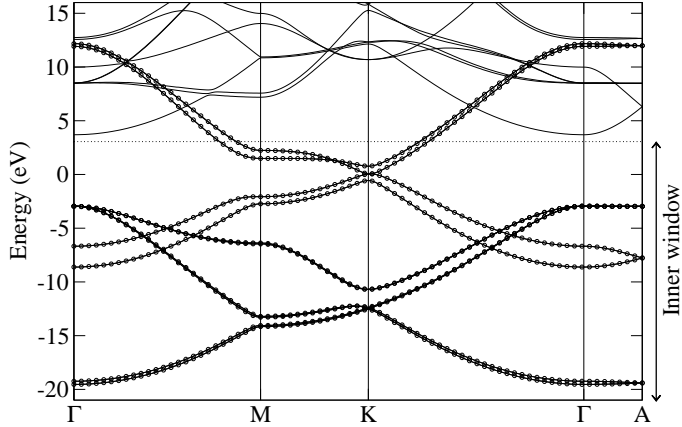}}
\caption{Band structure of graphite. Solid lines: original band structure
from a conventional first-principles calculation. Dotted lines:
Wannier-interpolated band structure. The zero of energy is the Fermi
level.}
\label{fig:graphite_bs} 
\end{center}
\end{figure}

As an example we consider the generation of MLWF to describe the states
around and below the Fermi level in Bernal (A-B-A) graphite. From the
bandstructure (Fig. \ref{fig:graphite_bs}) we expect that the minimum
number of Wannier functions needed to 
describe these states is 10 per unit cell (2.5 per atom). We shall see that
this choice corresponds to an intuitive chemical description of the system.

\begin{figure}[h]
\begin{center}
\scalebox{0.15}{\includegraphics{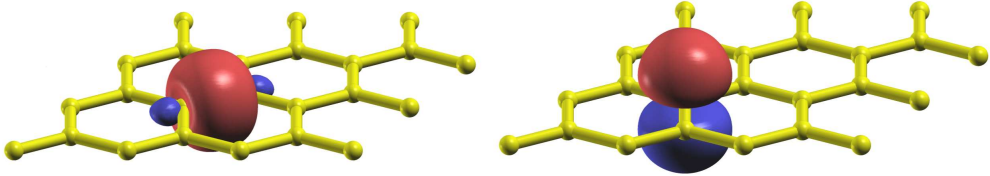}}
\caption{Isosurface contours of MLWF
  in graphite (red for positive value and blue for negative). [left]
  $\sigma$-type MLWF. [right] p$_z$-type MLWF.} \label{fig:graphite_wf} 
\end{center}
\end{figure}

 We perform the initial band-structure calculations using the \pwscf\
 code~\cite{pwscf}. A kinetic-energy cutoff of 30\,Ry is used for
 the plane-wave 
 expansion of the valence wavefunctions. The core-valence interaction is
 described by means of norm-conserving pseudopotentials in separable
 Kleinman-Bylander form~\cite{kleinman1}. We obtain the
 self-consistent ground state 
 using a 16$\times$16$\times$16 Monkhorst-Pack mesh of
 k-points~\cite{monkhorst} and a fictitious Fermi 
smearing~\cite{marzari97:_ensem_densit_funct_theor_ab} of 0.02\,Ry for
 the Brillouin-zone integration.  
  Then, we freeze the self-consistent potential and perform a
 non-self-consistent calculation on a uniform 6$\times$6$\times$6 grid of
 k-points. At each 
 k-point we calculate the first 20 bands. The required
 overlap matrices and projections are calculated using the
 post-processing routine  \texttt{pw2wannier90}, supplied with the
 \pwscf\ distribution. Projections onto atom centred sp$^2$ and p$_z$
 functions are used to construct the initial guess, and
\wannier\ is used to obtain the MLWF. The gauge-dependent and
gauge-independent spreads converge to machine precision in 300 and 70
steps, respectively. The resulting MLWF are a set of six symmetry-related 
bond-centred $\sigma$-orbitals and four atom-centred p$_z$-orbitals,
shown in Fig. \ref{fig:graphite_wf} 
(as plotted with the XCrySDen package~\cite{xcrysden}). Similar MLWF
were obtained for carbon nanotubes~\cite{lee05}. In
Fig.~\ref{fig:graphite_bs} we show the band structure of graphite
obtained using Wannier interpolation and compare it to the
band structure obtained from a full first-principles
calculation. Within the inner energy window the agreement is
essentially perfect.

\subsection{Lead}
Let us consider the generation of MLWF to describe the states
around and below the Fermi level in lead. 
As relativistic effects can be
significant in heavy elements, we include spin-orbit coupling
in our electronic structure calculation.
In lead, the 6s and 6p states
form an isolated set of bands around the Fermi level. The ground-state
structure of lead has both time-reversal and inversion symmetries,
so that each band is two-fold degenerate. We will obtain a set of eight
MLWF to describe these states.

 Once again we perform the initial band-structure calculations using
 the \pwscf\ code. A kinetic-energy cutoff of 45\,Ry is used for the
 plane-wave  expansion of the valence wavefunctions. The core-valence
 interaction is described by means of spin-orbit coupled
 norm-conserving pseudopotentials~\cite{corso05} in separable
 Kleinman-Bylander form. We obtain the self-consistent ground state 
 using a 16$\times$16$\times$16 Monkhorst-Pack mesh of k-points and
 a fictitious Fermi 
 smearing~\cite{marzari97:_ensem_densit_funct_theor_ab}  of 0.02\,Ry
 for the Brillouin-zone integration. 
 The self-consistent potential is frozen and we perform a
 non-self-consistent calculation on a uniform 12$\times$12$\times$12
 grid of k-points.  
 Then we calculate the required
 overlap matrices and projections using \texttt{pw2wannier90}. For the
 initial projection we use orbitals with sp$^3$ symmetry, four
 projections onto spin-up states and four onto spin-down states.

\begin{figure}[h]
\begin{center}
\scalebox{0.33}{\includegraphics{lead-band.eps}}\hspace*{0mm}
\scalebox{0.2}{\includegraphics{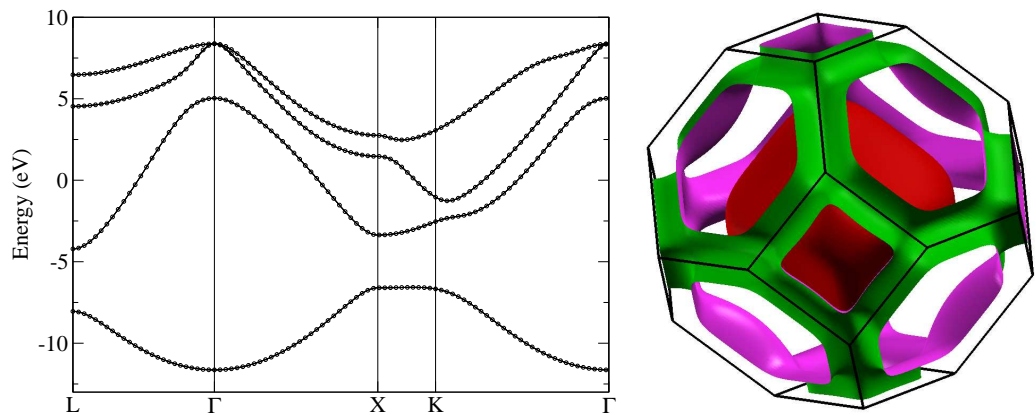}}
\caption{[left] Band structure of lead. Solid lines: original
  band structure from a conventional first-principles
  calculation. Dotted lines: Wannier-interpolated
  band structure. The zero of energy is the Fermi
  level. [right] Fermi surface of lead.} \label{fig:lead-bs}
\end{center}
\end{figure}

\wannier\ is used to obtain the MLWF. The gauge-dependent spread is
converged in 200 steps. In Fig.~\ref{fig:lead-bs} we show both the
band structure and Fermi surface obtained using Wannier
interpolation. Although the spin-orbit induced splitting is large at
certain points in the band structure (e.g., ~3\,eV at $\Gamma$), the
Fermi surface is not significantly different from a
scalar-relativistic calculation. 
The reader is referred to Ref.~\cite{xinjie06} for a detailed
discussion of how to construct spinor Wannier functions 
for ferromagnetic systems with broken time-reversal symmetry.

\section{Conclusions}

In conclusion, we present a new code called \wannier\ for computing
maximally-localised Wannier functions. \wannier\ is
freely available under the GNU General Public
Licence~\cite{GNU-GPL}. It is very user-friendly and is written in
Fortran90, employing modern programming techniques that enable the
addition of further functionality, such as transport properties or
interpolation of electron-phonon couplings, in an 
easy and modular fashion. \wannier\ has been seamlessly interfaced to
the \pwscf\ plane-wave DFT code~\cite{pwscf} and, at the time of
writing, interfaces to other codes, e.g.,
\textsc{abinit}~\cite{abinit}, \textsc{castep}~\cite{newcastep} and 
\textsc{fleur}~\cite{fleur}, are in progress. We hope that the
availability of \wannier\ will
encourage the wider use of maximally-localised Wannier functions in the
electronic structure community.

\begin{ack}
We would like to thank Stefano de
Gironcoli, Malgorzata Wierzbowska and Maria Peressi 
for the interface to the \pwscf\ first-principles
code~\cite{pwscf}. This work was funded by DARPA-PROM and by the
Laboratory Directed Research and Development Program of Lawrence
Berkeley National Laboratory under the Department of Energy Contract
No. DE-AC02-05CH11231.
\end{ack}

\appendix

%%%%%%%%%%%%%%%%%%%%%%%%%
\section{Sample files}\label{app:testrun}
%%%%%%%%%%%%%%%%%%%%%%%%%

Input file:

\begin{verbatim}

num_bands         =    20       
num_wann          =    10 
dis_win_max       =  19.2
dis_froz_max      =   9.8
dis_num_iter      =   400
num_iter          =   100

Begin Atoms_Frac
  C1    0.0000000000    0.0000000000    0.7500000000
  C1    0.3333333333    0.6666666667    0.2500000000
  C2    0.0000000000    0.0000000000    0.2500000000
  C2   -0.3333333333   -0.6666666667    0.7500000000
End Atoms_Frac

Begin Unit_Cell_Cart
  2.1304215583   -1.2299994602    0.0000000000
  0.0000000000    2.4599989204    0.0000000000
  0.0000000000    0.0000000000    6.8000000000
End Unit_Cell_Cart

Begin Projections     
  C1:sp2;pz
  C2:pz
End Projections       
    
bands_plot = true

begin kpoint_path
  G  0.0000   0.0000   0.0000    M  0.5000  -0.5000   0.0000
  M  0.5000  -0.5000   0.0000    K  0.6667  -0.3333   0.0000
  K  0.6667  -0.3333   0.0000    G  0.0000   0.0000   0.0000
  G  0.0000   0.0000   0.0000    A  0.0000   0.0000   0.5000
end kpoint_path

mp_grid      = 6 6 6  

begin kpoints
  0.00000000  0.00000000  0.00000000 
  0.00000000  0.00000000  0.16666667 
  0.00000000  0.00000000  0.33333333 
  0.00000000  0.00000000  0.50000000 
  0.00000000  0.00000000  0.66666667 
  0.00000000  0.00000000  0.83333333 
  0.00000000  0.16666667  0.00000000 
                 .
                 .
                 .
  0.83333333  0.83333333  0.66666667 
  0.83333333  0.83333333  0.83333333 
End Kpoints

\end{verbatim}

Output file (truncated):

\footnotesize
\begin{verbatim}
 ------------------------------------------------------------------------------
 Final State
  WF centre and spread    1  ( -0.354954,  0.614798,  5.100000 )     0.59341185
  WF centre and spread    2  ( -0.354954, -0.614798,  5.100000 )     0.59341185
  WF centre and spread    3  (  0.709907,  0.000000,  5.100000 )     0.59341190
  WF centre and spread    4  (  0.000000,  0.000000,  5.100000 )     1.04236015
  WF centre and spread    5  (  0.354954,  1.845201,  1.700000 )     0.59341120
  WF centre and spread    6  (  0.354954,  0.614798,  1.700000 )     0.59341085
  WF centre and spread    7  (  1.420514,  1.230000,  1.700000 )     0.59341100
  WF centre and spread    8  (  0.710140,  1.229999,  1.700000 )     1.04371374
  WF centre and spread    9  (  0.000000,  0.000000,  1.700000 )     1.04233128
  WF centre and spread   10  ( -0.710140, -1.229999,  5.100000 )     1.04368702
  Sum of centres and spreads (  2.130421,  3.689998, 33.999999 )     7.73256084
 
         Spreads (Ang^2)       Omega I      =     6.121019675
        ================       Omega D      =     0.032181361
                               Omega OD     =     1.579353494
    Final Spread (Ang^2)       Omega Total  =     7.732554530
 ------------------------------------------------------------------------------
\end{verbatim}

%%\bibliography{bib}

\begin{thebibliography}{10}

\bibitem{marzari97:_maxim_wannier}
Marzari, N. and Vanderbilt, D.,
\newblock Phys. Rev. B {\bf 56} (1997) 12847.

\bibitem{souza01:_maxim_wannier}
Souza, I., Marzari, N., and Vanderbilt, D.,
\newblock Phys. Rev. B {\bf 65} (2001) 035109.

\bibitem{marzari97:_ensem_densit_funct_theor_ab}
Marzari, N., Vanderbilt, D., and Payne, M.~C.,
\newblock Phys. Rev. Lett. {\bf 79} (1997) 1337.

\bibitem{cpmd}
See {\tt http://www.cpmd.org/}.

\bibitem{PhysRevB.59.9703}
Silvestrelli, P.~L.,
\newblock Phys. Rev. B {\bf 59} (1999) 9703.

\bibitem{mno-posternak}
Posternak, M., Baldereschi, A., Massidda, S., and Marzari, N.,
\newblock Phys. Rev. B {\bf 65} (2002) 184422.

\bibitem{pwscf}
Baroni, S. et~al.,
\newblock {\tt http://www.quantum-espresso.org/}.

\bibitem{PhysRevLett.87.246406}
Williamson, A.~J., Hood, R.~Q., and Grossman, J.~C.,
\newblock Phys. Rev. Lett. {\bf 87} (2001) 246406.

\bibitem{PhysRevB.67.085204}
Whittaker, D.~M. and Croucher, M.~P.,
\newblock Phys. Rev. B {\bf 67} (2003) 085204.

\bibitem{SMHB05}
Schillinger, M., Mingaleev, S.~F., Hermann, D., and Busch, K.,
\newblock Highly localized {W}annier functions for the efficient modeling of
  photonic crystal circuits,
\newblock in {\em Proceedings of SPIE: Volume 5733 -- Photonic Crystal
  Materials and Devices III}, edited by Adibi, A., Lin, S.-Y., and Scherer, A.,
  pages 324--335, Bellingham, WA, 2005, SPIE.

\bibitem{stengel07}
Stengel, M. and Spaldin, N.~A.,
\newblock Phys. Rev. B {\bf 75} (2007) 205121.

\bibitem{xinjie06}
Wang, X., Yates, J.~R., Souza, I., and Vanderbilt, D.,
\newblock Phys. Rev. B {\bf 74} (2006) 195118.

\bibitem{giustino07}
Giustino, F., Yates, J.~R., Souza, I., Cohen, M.~L., and Louie, S.~G.,
\newblock Phys. Rev. Lett. {\bf 98} (2007) 047005.

\bibitem{lee05}
Lee, Y.-S., Nardelli, M.~B., and Marzari, N.,
\newblock Phys. Rev. Lett. {\bf 95} (2005) 076804.

\bibitem{mostofi-dna}
Mostofi, A.~A. and Marzari, N.,
\newblock in preparation.

\bibitem{PhysRevLett.89.167204}
Ku, W., Rosner, H., Pickett, W.~E., and Scalettar, R.~T.,
\newblock Phys. Rev. Lett. {\bf 89} (2002) 167204.

\bibitem{anisimov:075125}
Anisimov, V.~I. and Kozhevnikov, A.~V.,
\newblock Phys. Rev. B {\bf 72} (2005) 075125.

\bibitem{lechermann:125120}
Lechermann, F. et~al.,
\newblock Phys. Rev. B {\bf 74} (2006) 125120.

\bibitem{PhysRevB.62.R16219}
Andersen, O.~K. and Saha-Dasgupta, T.,
\newblock Phys. Rev. B {\bf 62} (2000) R16219.

\bibitem{ChemPhysChem.6.1934.2005}
Zurek, E., Jepsen, O., and Andersen, O.~K.,
\newblock Chem. Phys. Chem. {\bf 6} (2005) 1934.

\bibitem{blount62}
Blount, E.~I.,
\newblock Formalisms of band theory,
\newblock in {\em Solid State Physics}, edited by Seitz, F. and Turnbull, D.,
  volume~13, pages 305--373, Academic Press, New York, 1962.

\bibitem{monkhorst}
Monkhorst, H. and Pack, J.,
\newblock Phys. Rev. B {\bf 13} (1976) 5188.

\bibitem{thygesen:026405}
Thygesen, K.~S., Hansen, L.~B., and Jacobsen, K.~W.,
\newblock Phys. Rev. Lett. {\bf 94} (2005) 026405.

\bibitem{Brouder:PRL.98.046402}
Brouder, C., Panati, G., Calandra, M., Mourougane, C., and Marzari, N.,
\newblock Phys. Rev. Lett. {\bf 98} (2007) 046402.

\bibitem{wanint}
Yates, J.~R., Wang, X., Souza, I., and Vanderbilt, D.,
\newblock Phys. Rev. B {\bf 75} (2007) 195121.

\bibitem{slater-koster}
Slater, J.~C. and Koster, G.~F.,
\newblock Phys. Rev. {\bf 94} (1954) 1498.

\bibitem{Si_MLWFS_2007}
Yates, J.~R., Mostofi, A.~A., Souza, I., and Marzari, N.,
\newblock in preparation.

\bibitem{SVD}
Golub, G.~H. and Loan, C. F.~V.,
\newblock {\em Matrix Computations}, chapter~2,
\newblock John Hopkins, New York, 3$^{\rm rd}$ edition, 1996.

\bibitem{stengel06b}
Stengel, M. and Spaldin, N.~A.,
\newblock Phys. Rev. B {\bf 73} (2006) 075121.

\bibitem{gygi03}
Gygi, F., Fattebert, J.-L., and Schwegler, E.,
\newblock Comp. Phys. Commun. {\bf 155} (2003) 1.

\bibitem{kleinman1}
Kleinman, L. and Bylander, D.,
\newblock Phys. Rev. Lett. {\bf 48} (1982) 1425.

\bibitem{vanderbilt1}
Vanderbilt, D.,
\newblock Phys. Rev. B {\bf 41} (1990) 7892.

\bibitem{xcrysden}
Kokalj, A.,
\newblock Comp. Mater. Sci. {\bf 28} (2003) 155.

\bibitem{corso05}
Corso, A.~D. and Conte, A.~M.,
\newblock Phys. Rev. B {\bf 71} (2005) 115106.

\bibitem{GNU-GPL}
See {\tt http://www.gnu.org/} for details.

\bibitem{abinit}
Gonze, X. et~al.,
\newblock Comp. Mater. Sci. {\bf 25} (2002) 478.

\bibitem{newcastep}
Clark, S.~J. et~al.,
\newblock Zeitschrift f{\"{u}r} Kristallographie {\bf 220} (2005) 567.

\bibitem{fleur}
See {\tt http://www.flapw.de/}.

\end{thebibliography}

\end{document}